\documentclass[floatfix,aps,twocolumn,prl,superscriptaddress]{revtex4}
\usepackage{amsmath}
\usepackage{amssymb}
\usepackage{graphicx}
\usepackage{dcolumn}
\usepackage{natbib}
\usepackage{bm}
\usepackage{epstopdf}
\usepackage{multirow}
\begin{document}
\title{Electronic correlations and unconventional spectral weight transfer in BaFe$_{2-x}$Co$_{x}$As$_{2}$}
\author{A. A. Schafgans}
\email{aschafgans@physics.ucsd.edu}
\author{S. J. Moon}
\author{B. C. Pursley}
\author{A. D. LaForge}
\affiliation{Department of Physics, University of California, San Diego, La Jolla, California 92093, USA}
\author{M. M. Qazilbash}
\affiliation{Department of Physics, College of William and Mary, Williamsburg, VA 23187, USA}
\author{A. S. Sefat}
\affiliation{Materials Science and Technology Division, Oak Ridge National Laboratory, Oak Ridge, Tennessee 37831, USA}
\author{D. Mandrus}
\affiliation{Materials Science and Technology Division, Oak Ridge National Laboratory, Oak Ridge, Tennessee 37831, USA}
\affiliation{Department of Materials Science and Engineering, University of Tennessee, Knoxville, TN 37996}
\author{K. Haule}
\author{G. Kotliar}
\affiliation{Department of Physics and Astronomy, Rutgers University, Piscataway, New Jersey 08854, USA}
\author{D. N. Basov}
\affiliation{Department of Physics, University of California, San Diego, La Jolla, California 92093, USA}
\date{\today}

\begin{abstract}
We report an infrared optical study of the pnictide high-temperature superconductor BaFe$_{1.84}$Co$_{0.16}$As$_{2}$ and its parent compound BaFe$_{2}$As$_{2}$. We demonstrate that electronic correlations are moderately strong and do not change across the spin-density wave transition or with doping. By examining the energy scale and direction of spectral weight transfer, we argue that Hund's coupling \emph{J} is the primary mechanism that gives rise to correlations.
\end{abstract}

\maketitle

Strongly correlated materials present some of the most interesting challenges to both experimental and theoretical physicists alike. A profound experimental manifestation of correlations is the renormalization of electronic bands due to interactions that are not included in the standard band theory description \protect\cite{Millis-PRB72-224517-2005,Qazilbash-NatPhys5-647-2009,BasovRMP}. Correlations are well known to give rise to complex phase diagrams, as evidenced by the long-standing problem of high-temperature superconductivity in cuprates \protect\cite{Basov-RMP77-721-2005,Comanac-NatPhys4-287-2008}. Since the discovery of the iron-pnictide superconductors, one pressing question in this field is whether correlations in general, and Mott physics specifically, are among necessary preconditions for the high-$T_c$ phenomenon. While both the cuprates and pnictides exhibit an antiferromagnetically ordered ground state in proximity with the superconducting phase (\protect\cite{Paglione-NatPhys6-645-2010,Si-PRL101-076401-2008} and references therein), the parent compounds of the iron-pnictide materials are metallic with striped spin-density wave order, unlike the antiferromagnetic Mott insulating cuprate parent compounds. As we will show for the 122 class of pnictides, the influence of magnetism via Hund's coupling is the predominant correlation mechanism and is crucial to the electronic properties of the paramagnetic normal and superconducting states. 
\begin{figure}
\centering
\includegraphics[width=3.4in]{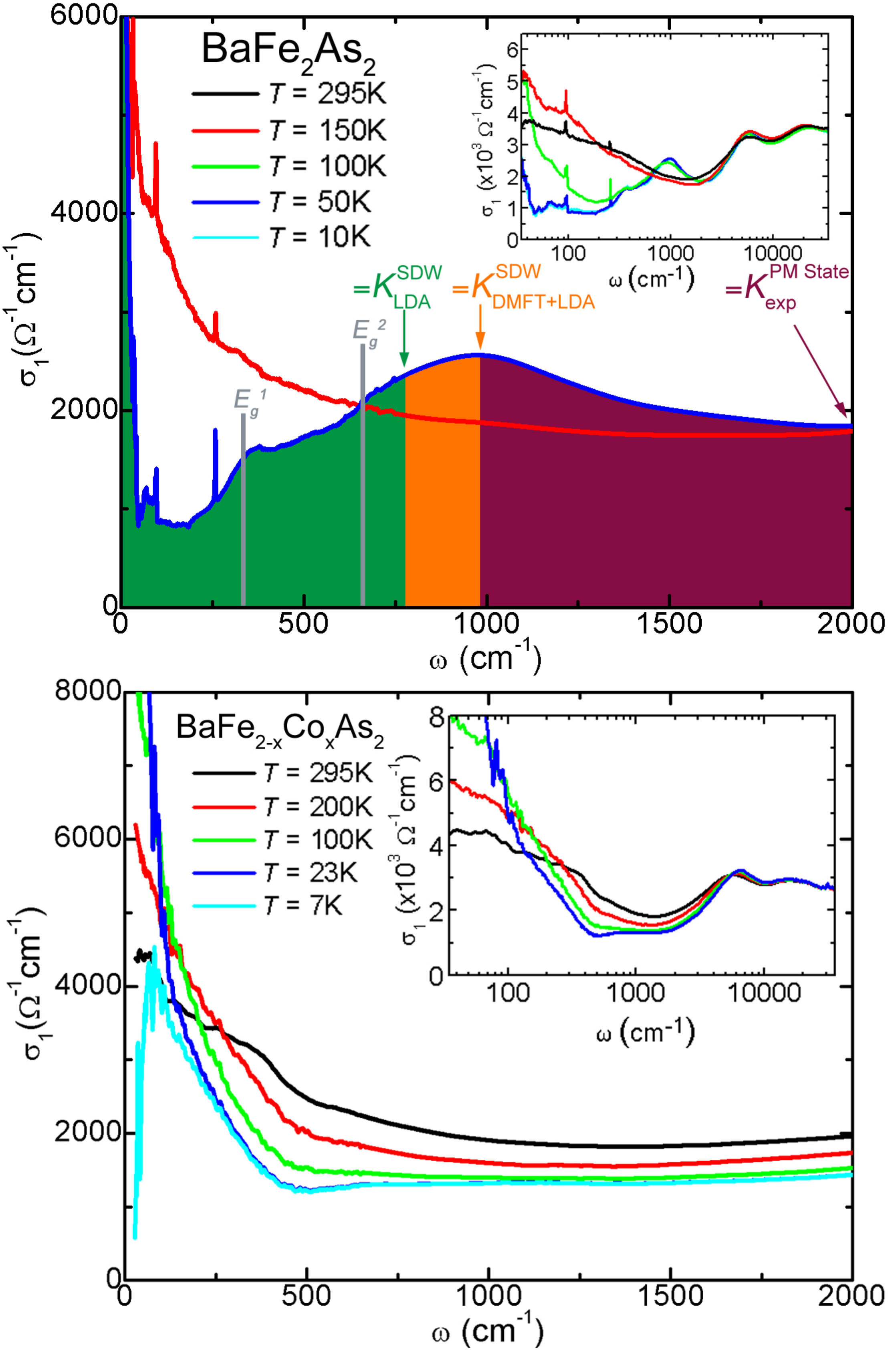}
\caption{Real part of the optical conductivity $\sigma_1$ in Ba122 (Top) in the PM ($T=150K$) and SDW ($T=50K$) state and Co-Ba122 (Bottom) in the PM normal state and superconducting state ($T=7K$). For Ba122, we illustrate the necessary integration cutoffs $\Omega$ = 777, 969, and 2000 cm$^{-1}$ in order to recover the predicted LDA, LDA+DMFT, and experimentally observed PM state kinetic energy values, respectively. The third cutoff demonstrates that the spectral weight transfer due to the SDW gaps extends to 2000 $cm^{-1}$ and not above.}
\end{figure} 

We measured the optical reflectivity of the parent BaFe$_{2}$As$_{2}$ (Ba122) and optimally doped BaFe$_{1.84}$Co$_{0.16}$As$_{2}$ (Co-Ba122) compounds as a function of temperature over a broad spectral range and extracted the optical constants as described in the supplementary materials (SM). The real part of the optical conductivity $\sigma_1(\omega)$ is plotted in Figure 1. The following discussion will distinguish properties of the spin-density wave (SDW) state, which applies only to the parent compound for $T<T_{SDW}\approx $140K, from properties of the paramagnetic normal state (PM), which applies to $T>T_{SDW}$ in Ba122 and $T>T_c$ in Co-Ba122. All of the $\sigma_1(\omega)$ spectra in the PM state involve a broad Drude response extending to at least 1500 cm$^{-1}$. In the SDW state, $\sigma_1(\omega)$ is dominated at low energy by the formation of the SDW gaps, exhibiting only a very narrow Drude response \protect\cite{Hu-PRL101-257005-2008,Moon-PRB81-205114-2010}. Doping does not seem to drastically alter $\sigma_1(\omega)$ in the PM state.

The renormalization of electronic bands due to correlations has a significant effect on two specific observables in infrared spectroscopy: one is a reduced oscillator strength of the Drude response compared with band theory predictions and the other is the energy scale of spectral weight transfer. We now explore each of these observables in turn. (In the SM, we present more details about the samples, experiment, and determination of the SDW gap values.)

In order to quantify the first observable, a reduction of the Drude response compared to band theory, we use a truncated version of the \emph{f}-sum rule \protect\cite{Millis-PRB72-224517-2005}. The experimental kinetic energy $K_{exp}$ is proportional to the spectral weight of the Drude component of the optical response and is determined via an integral of the real part of the optical conductivity up to a cut-off value $\Omega$:
\begin{equation}
{K_{exp}(\Omega)} = {{120\over{\pi}}\int_{0}^{\Omega}\sigma_1(\omega)d\omega}.
\end{equation}
Here we have formulated the equation for kinetic energy in units of cm$^{-2}$. The cut-off value $\Omega$ should be high enough so as to account for all of the Drude weight, yet not so high as to include significant contributions from interband transitions. We then normalize $K_{exp}$ to $K_{LDA}$, the band theory estimate under the local density approximation (LDA). The same procedure was used to explore correlations in several different families of exotic superconductors including the pnictides \protect\cite{Qazilbash-NatPhys5-647-2009,BasovRMP,Basov-Chubukov}. The ratio spans from the extremely correlated case of a fully localized Mott-insulator (eg. the cuprate parent compounds where $K_{exp}/K_{LDA} \approx 0$) to electronically uncorrelated materials such as a fully itinerant metal (eg. copper where $K_{exp}/K_{LDA} \approx 1$). The comparison is instructive because band theory with LDA does not take into account Coulomb repusion \emph{U} or magnetic interactions that can renormalize the electronic bandwidth and consequently reduce the kinetic energy. Therefore, the ratio of $K_{exp}/K_{LDA}$ emphasizes the significance of correlations due to processes beyond the band theory description.

In Figure 2a-c, we plot $K_{exp}/K_{LDA}$ as a function of the cut-off frequency, $\Omega$. The LDA kinetic energy values are given in the caption and were obtained as described in a recent work by Yin, \emph{et. al.} \cite{Yin-arXiv1007.2867}. Using arrows, we point out the various cut-offs explained in the SM. Irrespective of the cut-off criteria, we find that for Ba122, $K_{exp}/K_{LDA}$ = 0.25-0.29 in the PM and 0.2-0.34 in the SDW state. In Co-Ba122, we observe a similar value in the PM normal state: $K_{exp}/K_{LDA}$ = 0.15-0.31. Our results  demonstrate that correlations are present in the PM state and remain unchanged at low temperature in the fully-developed SDW state and with doping into the superconducting state.
\begin{figure}
\centering
\includegraphics[width=3.4in]{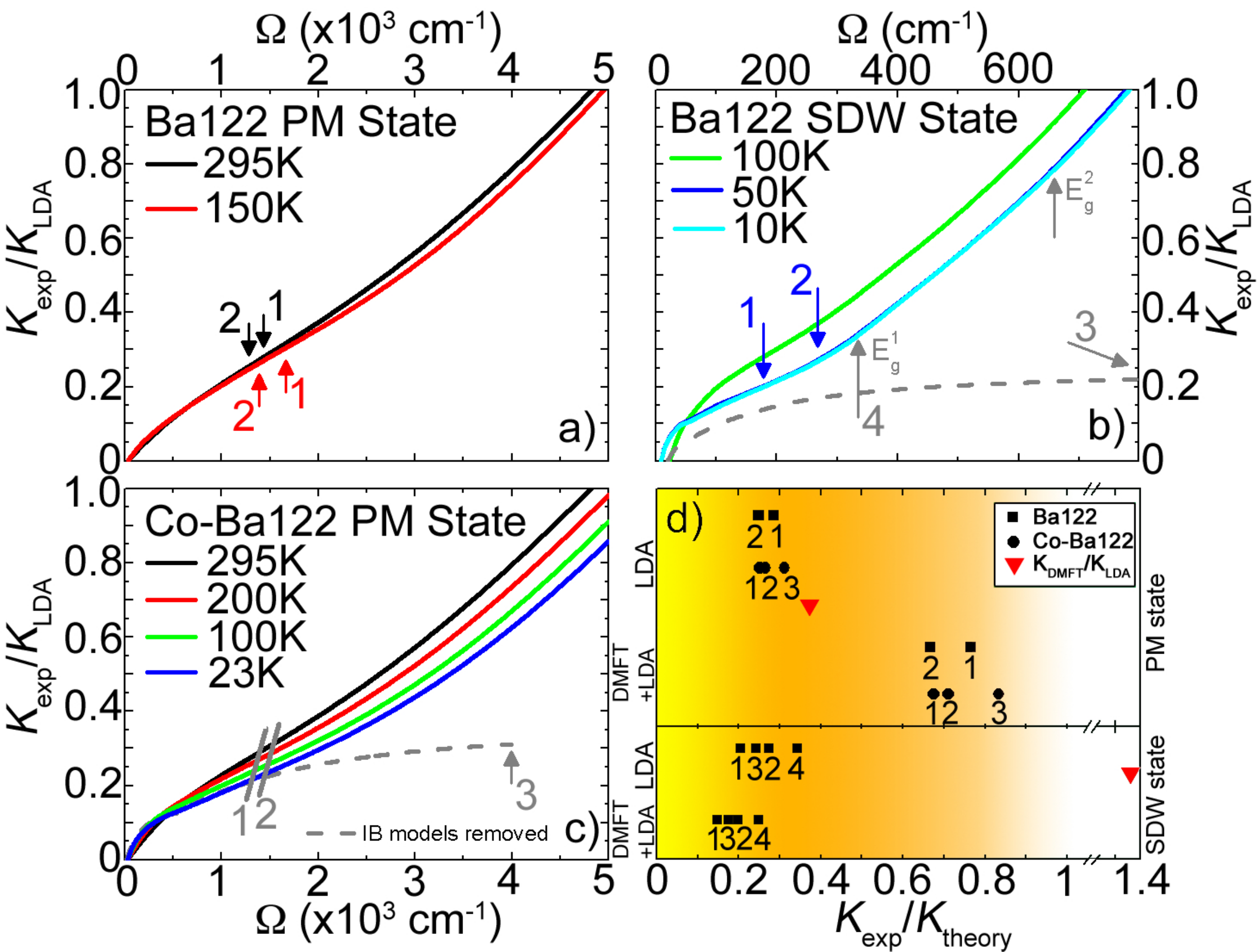}
\caption{Panels A-C: The kinetic energy ratio $K_{exp}(\Omega)/K_{LDA}$ in Ba122 and Co-Ba122, as a function of integration cutoff value $\Omega$ (Eq. 1). The ratio for Ba122 in the PM (panel a) and SDW (panel b) states, and for Co-Ba122 in the normal PM state (panel c). The numbers and arrows indicate the results of the four methods for determining the kinetic energy integration cutoff, described in the SM. Panel D: The temperature averaged results of the kinetic energy ratio $K_{exp}(\Omega)/K_{theory}$ in the PM and SDW states, compared with both LDA and LDA+DMFT results, including the theoretical kinetic energy values. The triangles show the value of the ratio of the two theoretical kinetic energies determined via LDA and LDA+DMFT in the PM and SDW state. The values from theory are: $K_{LDA}^{PM}$= (21133 cm-1)$^2$, $K_{LDA}^{SDW}$= (6937 cm-1)$^2$, $K_{DMFT}^{PM}$= (12905 cm-1)$^2$, and $K_{DMFT}^{SDW}$= (8147 cm-1)$^2$.}
\end{figure}

One recent attempt to theoretically explain the optical conductivity and magnetic moment in the PM state of Ba122 utilizes dynamical mean field theory in conjunction with the local density approximation (LDA+DMFT) \protect\cite{Yin-arXiv1007.2867}. In Fig. 2 (panel d), we have also plotted the ratio $K_{exp}/K_{DMFT}$, where $K_{DMFT}$ corresponds to the kinetic energy predicted using LDA+DMFT. This theory can be understood to describe a PM metal with a fluctuating magnetic moment. In Ba122, we find that in the PM state $K_{exp}/K_{DMFT}$ = 0.67-0.77 while in Co-Ba122 the ratio is $K_{exp}/K_{DMFT}$ = 0.68-0.83. We therefore confirm the ability of LDA+DMFT to more accurately describe the optical response in the PM state. However, this conclusion does not hold for the SDW state. In fact, LDA+DMFT underestimates the degree of correlations by an extra 30\% over LDA alone, giving $K_{exp}/K_{DMFT}$ = 0.15-0.25. This is surprising due to the qualitative agreement between our data and the theoretical specta. (We note that this discrepancy between $K_{exp}$ and $K_{DMFT}$ may be entirely explained by the excessively high background conductivity in the theoretical results \protect\cite{Yin-arXiv1007.2867}.)

Comparison with theory allows us to determine that the strength of correlations in BaFe$_{2-x}$Co$_{x}$As$_{2}$ is significant ($K_{exp}(\Omega)/K_{LDA}\approx 0.4$, see SM): the strength is on par with what is known to produce the highest $T_c$ in optimally doped cuprates \protect\cite{Millis-PRB72-224517-2005,Qazilbash-NatPhys5-647-2009,Basov-Chubukov}. Such reduced kinetic energy from theoretical expectations implies that additional spectral weight (SW) is present at energies higher than the Drude response. In order to elucidate the mechanisms that drive correlations in the Ba122 pnictides, one must determine the energy scale associated with this SW. However, the higher energy SW contribution is not immediately evident in $\sigma_1(\omega)$ due to the obscuring presence of interband transitions. Similar complications arise with doped Mott insulators where the energy scale of \emph{U} overlaps with the interband structure. Nevertheless, it is possible to gain direct information about the magnitude of \emph{U} in doped Mott insulators by carefully monitoring the distribution of SW as it changes with doping and temperature \protect\cite{Qazilbash-PRB77-115121-2008,Stewart-PRB83-075125-2011}. We utilize this approach in the pnictides by monitoring the energy scale of SW transfer in $\sigma_1(\omega)$.

In Figure 3, we plot the ratio of the integrated SW at low temperature $K_{exp}(T,\Omega)$ to the room temperature value $K_{exp}(295K, \Omega)$, where the cut-off value $\Omega$ extends throughout the entire measured frequency range. This ratio emphasizes the relevant energy scale of SW transfer. If there is a transfer of SW from high to low energy, the SW ratio will exceed 1 at low energy and then smoothly approach 1 until the full energy scale of the Drude oscillator is reached. If there is a transfer of SW from low to high energy, the SW ratio will fall below 1 until the total energy scale of SW transfer is reached.
\begin{figure}
\centering
\includegraphics[width=3.4in]{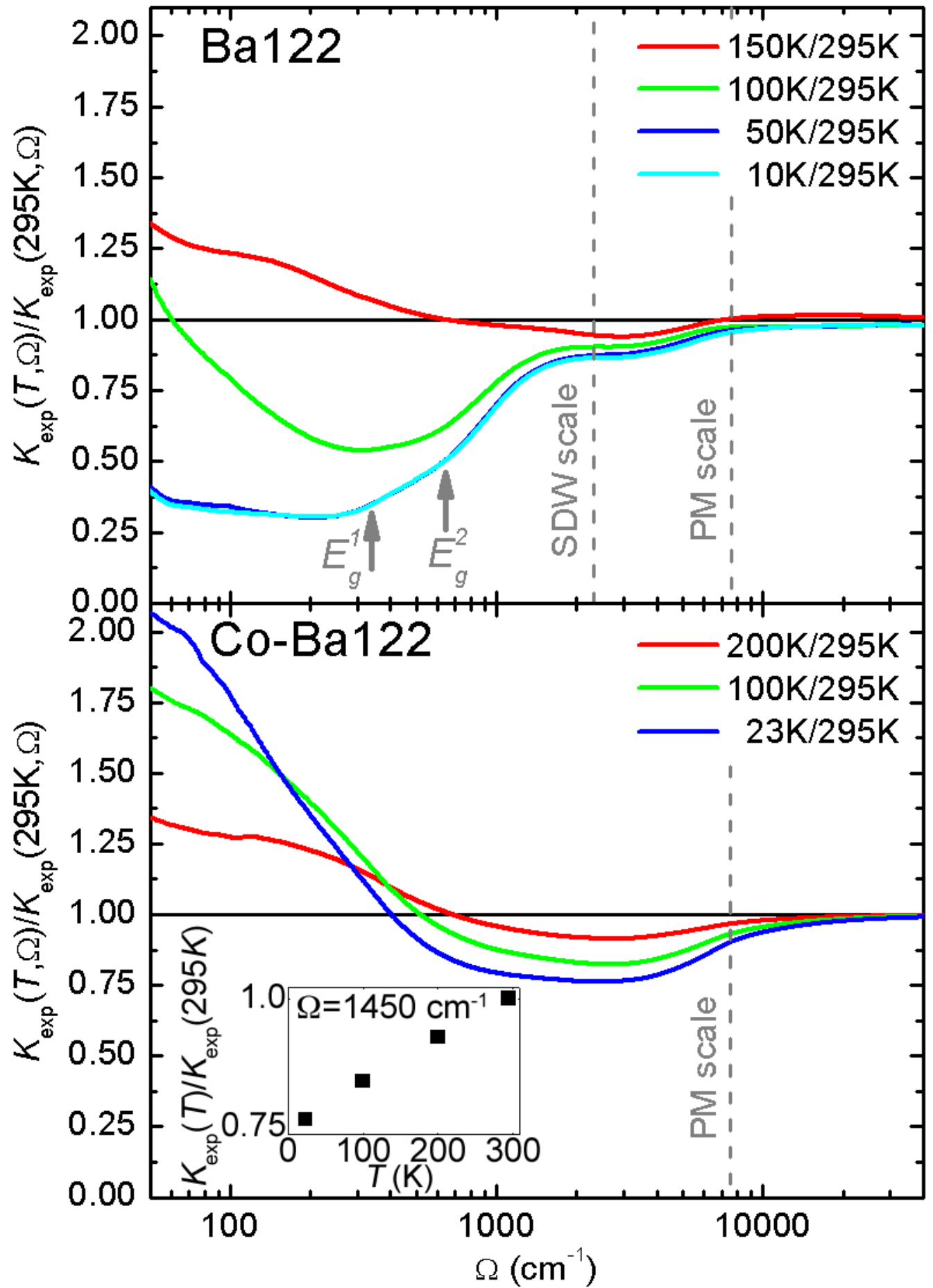}
\caption{Ratio of the integrated conductivity $K_{exp}(T,\Omega)/K_{exp}(295 K, \Omega)$ as a function of cutoff value $\Omega$ and temperature over the entire measured frequency range in Ba122 (top panel) and Co-Ba122 (bottom panel). The ratios are taken with respect to the $T=295$K data in order to display the temperature evolution of the PM and SDW SW transfer. The two vertical dashed bars show the energy scale below which all of the SW transfer takes place, for both the PM and SDW states. The energy scale of SW transfer in the PM state is $\approx 8000$ cm$^{-1}$ and in the SDW state is $\approx 2000$ cm$^{-1}$. Inset: Temperature dependence of the kinetic energy of the Drude response, normalized to the room temperature value, at a cut-off $\Omega$=1450cm$^{-1}$.}
\end{figure}

We now discuss the SW ratios in the PM state (Fig. 3). As temperature decreases, the Drude response narrows and some SW shifts lower in energy. This is in accordance with expectations for a metal and causes the SW ratio to exceed 1 below $\approx$ 600 cm$^{-1}$. Yet between $\approx$ 600 cm$^{-1}$ - 8000 cm$^{-1}$, the ratios decrease to well below 1. There is a minimum in the PM state ratios near 3000 cm$^{-1}$, indicating a turning point in SW transfer: below this value, SW is depleted while above this value, SW is amassed. By $\approx$8000 cm$^{-1}$, the ratios return to $\approx$ 1 as all of the SW is recovered. Therefore, SW is transferred from the Drude response into the region between 3000 - 8000 cm$^{-1}$ in both Ba122 and Co-Ba122. The energy scale of SW transfer is determined by observing the energy range over which most of the spectral weight is recovered, marked in Fig. 3 with vertical dashed lines. By this criterion, the energy scale for correlations in the PM state is $\approx$8000 cm$^{-1}$. 

In the SDW state in Ba122, we observe the superposition of two concomitant processes each with its own energy scale of SW transfer. The energy scale relevant for correlations in the PM state persists in the SDW state. In addition, there is SW transfer at lower energies, consistent with the formation of the SDW gaps. When the SDW state forms, a peak in $\sigma_1(\omega)$ emerges at finite frequency. Therefore, the SW in the remaining Drude response plus the SW in the SDW peak should be equal to the SW in the Drude response in the PM state. This is the case in Ba122, exemplified in Fig. 1 (top panel) where we show that in order to recover SW equal to the Drude response in the PM state, the integration cut-off must be equal to the frequency where the SDW peak returns to the PM state value ($\Omega \approx$ 2000 cm$^{-1}$). Therefore, we contend that the SW transfer in the SDW state can be understood as a superposition of two processes: the gapping of states due to SDW order, restricted to energies below $\approx 2000$ cm$^{-1}$, and SW transfer from low to high energy over the same energy scale as in the PM state, extending to energies as high as $\approx 8000$ cm$^{-1}$. This shows that the SDW state does not introduce additional electronic correlations in Ba122.

Having determined the energy scale of SW transfer, we are now in a position to compare our observations with the energy scales potentially relevant for correlations in the 122 materials: the Hubbard \emph{U} and Hund's coupling \emph{J}. Experimentally, x-ray absorption spectroscopy (XAS) can directly access the magnitude of \emph{U} and \emph{J}. A recent work determined that in Ba122, $U\leq$ 2eV $\approx$16000 cm$^{-1}$ and $J\approx$ 0.8eV $\approx$6400 cm$^{-1}$ \protect\cite{Yang-PRB80-014508-2009} while other XAS experiments determined, for a broad class of pnictides, that \emph{U}=3-4eV$\approx$24000-32000cm$^{-1}$ and $J\approx$ 0.8eV $\approx$6400 cm$^{-1}$, in agreement with constrained DFT calculations (\protect\cite{Anisimov-PhysicaC469-442-2009} and references therein). In LDA+DMFT calculations, including those that produce an optical conductivity closely resembling our data \cite{Yin-arXiv1007.2867}, the interaction strengths were \emph{U} = 3 - 5 eV and \emph{J} = 0.6 - 0.9 eV $\approx$ 3,600 - 7,200 cm$^{-1}$ \protect\cite{Anisimov-PhysicaC469-442-2009,Zhou-PRL105-096401-2009,Kutepov-PRB82-045105-2010}. On the other hand, mean field theory results suggest an intermediate value of \emph{U}=2-4 eV \cite{Yu-PRB79-104510-2009}, while a multiorbital Hubbard model under the mean field approximation gives \emph{U}$\approx$1.5 eV and $J\approx$0.3-0.6 eV \protect\cite{Luo-PRB82-104508-2010}. 

All of these results point to a picture where $U\approx$2-5eV while \emph{J} is some fraction thereof: $J\approx$0.6-0.9 eV. Comparing the values of \emph{U} and \emph{J} with our SW data, it is clear that the Coulomb scale \emph{U} $\approx$ 2-5 eV is larger than the observed SW transfer in our samples \protect\cite{CoulombNote}, especially considering the fact that the bulk of the literature we reviewed leans towards the larger value of \emph{U} within the above interval. Moreover, as demonstrated, SW does not significantly change with doping, while in a Mott insulator the effective number density of carriers increases linearly with doping. 

Of additional importance is the \emph{direction} of SW transfer. In strongly correlated metals derived from Mott insulators, the Drude oscillator strength grows at lowered temperatures due to SW transfer from high to low energy, over an energy scale of the Hubbard \emph{U}/2 \protect\cite{Kotliar-PhysToday57-March53-2004}. The in-plane response of other correlated metals such as V$_2$O$_3$ \protect\cite{Qazilbash-PRB77-115121-2008,footnote} and a host of CuO$_2$ - based materials across the doping-phase diagram \protect\cite{Nicoletti-arxiv1101.0745,Ortolani-PRL94-067002-2005,Molegraaf-Science295-2239-2002,Santander-PRL88-097005-2002}, also show SW transfer from high to low energy with decreasing temperature, over an energy scale associated with \emph{U}. Therefore, in systems where the Coulomb repulsion \emph{U} defines the correlation scale, SW transfer is from high to low energy as temperature is decreased, resulting in a larger Drude response at low temperature. The situation in the 122 pnictides is quite different, where the SW transfer at low temperature is from \emph{low to high} energy. As can be readily observed in the Fig. 3 inset, SW in the Drude response \emph{decreases} as temperature is lowered \protect\cite{Wu-PRB83-100503R-2011,Barisic}. Such behavior is reminiscent of doped semiconductors and sets the pnictides apart from correlated metals derived from Mott insulators.

By studying both the direction and energy scale of SW transfer, the evidence weighs in favor of a correlation mechanism predominantly due to Hund's coupling: \emph{J}$\approx$0.6-0.9 eV, which is equal to the scale of SW transfer that we observe. These results are unexpected in light of doped Mott-insulators such as the cuprates, where electronic correlations decrease as the material is doped. Moreover, unconventional SW transfer from low to high energy is also observed in P-doped and Co-overdoped Ba122, indicating that it may be a general phenomenon \protect\cite{soonjae-unpub}. In the future, a systematic study of detwinned samples may reveal important anisotropies \protect\cite{Dusza,Chu-Science329-824-2010} of the SW transfer.

Note added in proof: After completion of this work, Wang, \emph{et. al.} posted similar conclusions about the importance of Hund's coupling \protect\cite{Wang}.

This work was supported by the NSF 1005493 and the AFOSR. D. Mandrus was supported by the U.S. DOE. G. Kotliar was supported by NSF grant DMR-0906943.

\newpage

\section{Supplementary Materials}

\section{A. Samples and Experiment}

The samples in this study were square platelet single crystals of BaFe$_{2}$As$_{2}$ (Ba122) approximately 2 x 2 x 0.1 mm$^3$ in size, and Ba(Fe$_{2-x}$Co$_{x}$As)$_{2}$ (Co-Ba122) approximately 4 x 4 x 0.1 mm$^3$ in size. The doping of \emph{x}=0.08, measured via energy dispersive x-ray spectroscopy, places the superconducting sample just beyond optimal doping, with a corresponding \emph{T}$_{c}$= 22 K \protect\cite{Ning-arXiv:0811.1617,Sefat-arXiv:0807.2237,Sefat-PRB79-094508-2009}. In Ba122, two phase transitions have been observed as a function of temperature; a structural phase transition at $T_{STR}$=140K from the high temperature tetragonal (HTT) TrCr$_{2}$Si$_{2}$ type structure with paramagnetic (PM) order to a low temperature orthorhombic (LTO) lattice structure, and a magnetic phase transition with the onset of spin density wave (SDW) order at $T_{STR} \gtrsim T_{SDW} \approx$ 140 K \protect\cite{Sefat-arXiv:0807.2237,Hu-PRL101-257005-2008,Sefat-PRB79-094508-2009}. In the Co-Ba122 sample, the Co-concentration is sufficient to eliminate both the structural and magnetic phase transitions observed in Ba122. Fabrication and characterization are described elsewhere \protect\cite{Sefat-arXiv:0807.2237}. 

We measured near-normal incidence reflectance \emph{R}($\omega$) of the \emph{ab} face, over a frequency range of $\omega \approx$ 20 to 12,000 cm$^{-1}$. The reflectance measurements were performed for a variety of temperatures, ranging from \emph{T} = 6.5 K to 295 K, normalized using an in-situ gold overcoating technique \protect\cite{Homes}. Additionally, we performed variable-angle spectroscopic ellipsometry from $\omega \approx$ 5,500 to 45,000 cm$^{-1}$. In order to extract the optical constants, we performed a Kramers-Kronig constrained variational analysis using refFIT software, which utilizes a multi-oscillator fit of the reflectivity data anchored by the dielectric function measured through ellipsometry \protect\cite{Kuzmenko-RevSciInstrum76-083108-2005,Qazilbash-arXiv:0808.3748}. We also performed a Kramers-Kronig inversion of \emph{R}($\omega$) to extract the optical constants and obtained nearly identical results.

As shown in Fig. 1, in Ba122 for temperatures $T < T_{SDW}$, a significant loss of conductivity below $\approx$ 700 cm$^{-1}$ results in the development of a large mode at $\approx$ 1,000 cm$^{-1}$ where $\sigma_{1}(\omega)$ is greater than the PM state value extending into the MIR region. Additionally, there is a smaller mode that develops near 350 cm$^{-1}$. As has been done previously \protect\cite{Moon-PRB81-205114-2010,Hu-PRL101-257005-2008}, we attribute these features and the increase in mid-infrared conductivity to the redistribution of spectral weight from the region of the SDW gap. We determined the two SDW gap values from the local minima in the imaginary part of the optical conductivity ($\sigma_{2}(\omega)$, Fig. 1) \protect\cite{Li-NatPhys4-532-2008}: $E_{g}^{1}$ = 336 $\pm$ 3 cm$^{-1}$ and $E_{g}^{2}$ = 656 $\pm$ 2 cm$^{-1}$.   
\begin{figure}
\centering
\includegraphics[width=3.4in]{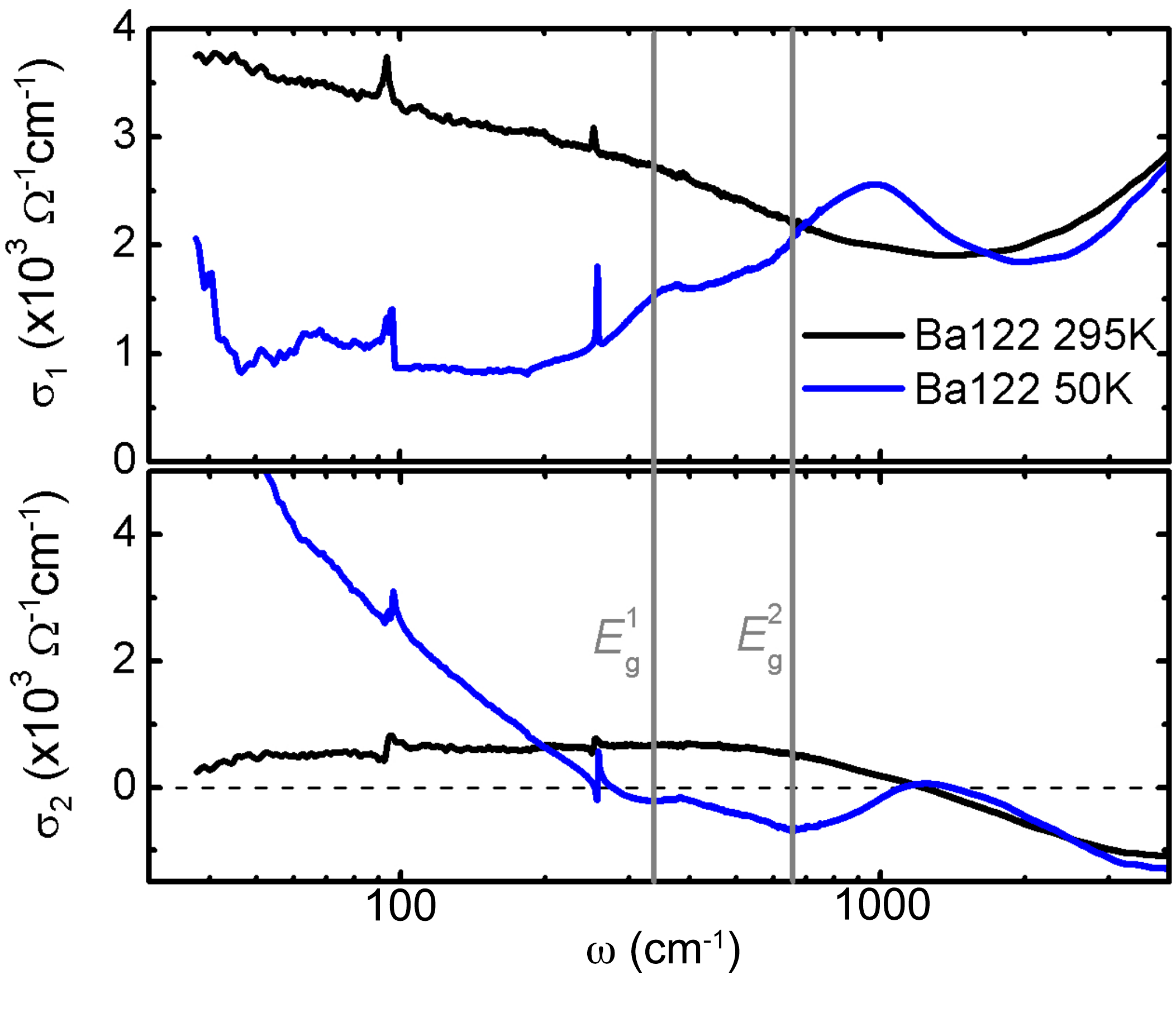}
\caption{Real and imaginary parts of the optical conductivity $\sigma_1(\omega)$ (top panel) and $\sigma_2(\omega)$ (bottom panel), respectively. Two representative temperatures are shown in the parent compound Ba122 to exemplify the differences between the PM and SDW states. The SDW gaps were determined by measuring the local minima in $\sigma_2(\omega)$, shown here with vertical bars.}
\end{figure}

\section{B. Determination of kinetic energy}

The experimental kinetic energy $K_{exp}$ is equal to the integral of the Drude response of the real part of the optical conductivity ($\sigma_1(\omega)$) (Eq. 1). In practice, this integral can be performed in two ways. One can integrate up to a frequency cut-off $\Omega_D$ that is high enough to encompass most of the Drude conductivity without incorporating too much interband (IB) conductivity. Alternatively, one can model the IB contribution, subtract the model from the $\sigma_1(\omega)$ data, and integrate the remainder that is presumably only due to the Drude response. We determined $K_{exp}$ using both methods. 

Two reasonable places for the integration cut-off $\Omega_D$ are based on estimates of the upper bound of the Drude energy scale before $\sigma_1(\omega)$ becomes dominated by IB contributions. One place is the minimum value of the conductivity, $\sigma_1^{min}(\omega)$; the point between the roll-off of the Drude response and the initial increase due to IB processes and is the most common cut-off value. A second cut-off value is given by the crossover of the imaginary part of the conductivity $\sigma_2(\omega)$ from positive to negative values (this can also be thought of as the crossover of the real part of the dielectric function $\epsilon_1(\omega)$ from negative to positive values). $\sigma_2(\omega)$ will be driven negative only by IB-type processes once the majority of the Drude response is exhausted. Of course, the actual frequency position of $\sigma_1^{min}(\omega)$ and $\sigma_2(\omega)$ will be determined by the relative spectral weight and energy separation of the Drude and IB processes. In the SDW state, we used the additional criterion of the lower SDW gap $E_g^1$ as our best estimate for $\Omega_D$.

As far as the IB model, for Co-Ba122 the presence of a large incoherent component of the conductivity in the PM state can be modeled as either a broad Drude or as a Lorentzian. We were guided by the spectral weight and kinetic energy analysis presented in our manuscript, as well as in the literature \protect\cite{Tu-arxiv1008.3098,Fang-PRB80-140508R-2009,Nakajima-PRB81-104528-2010,WuPRB81}, which demonstrate the majority, if not all of the incoherent spectral weight must be considered to be part of the Drude response. Therefore, the resultant IB model (Fig. 2) does not include the incoherent states that contribute to the broad Drude response. (We note that of all the methods, in the PM state there is the greatest agreement between the spectral weight found using the IB model for Co-Ba122 and the LDA + DMFT approach, while in the SDW state our best estimate uses $E_g^1$ as the integration cut-off.)
\begin{figure}
\centering
\includegraphics[width=3.4in]{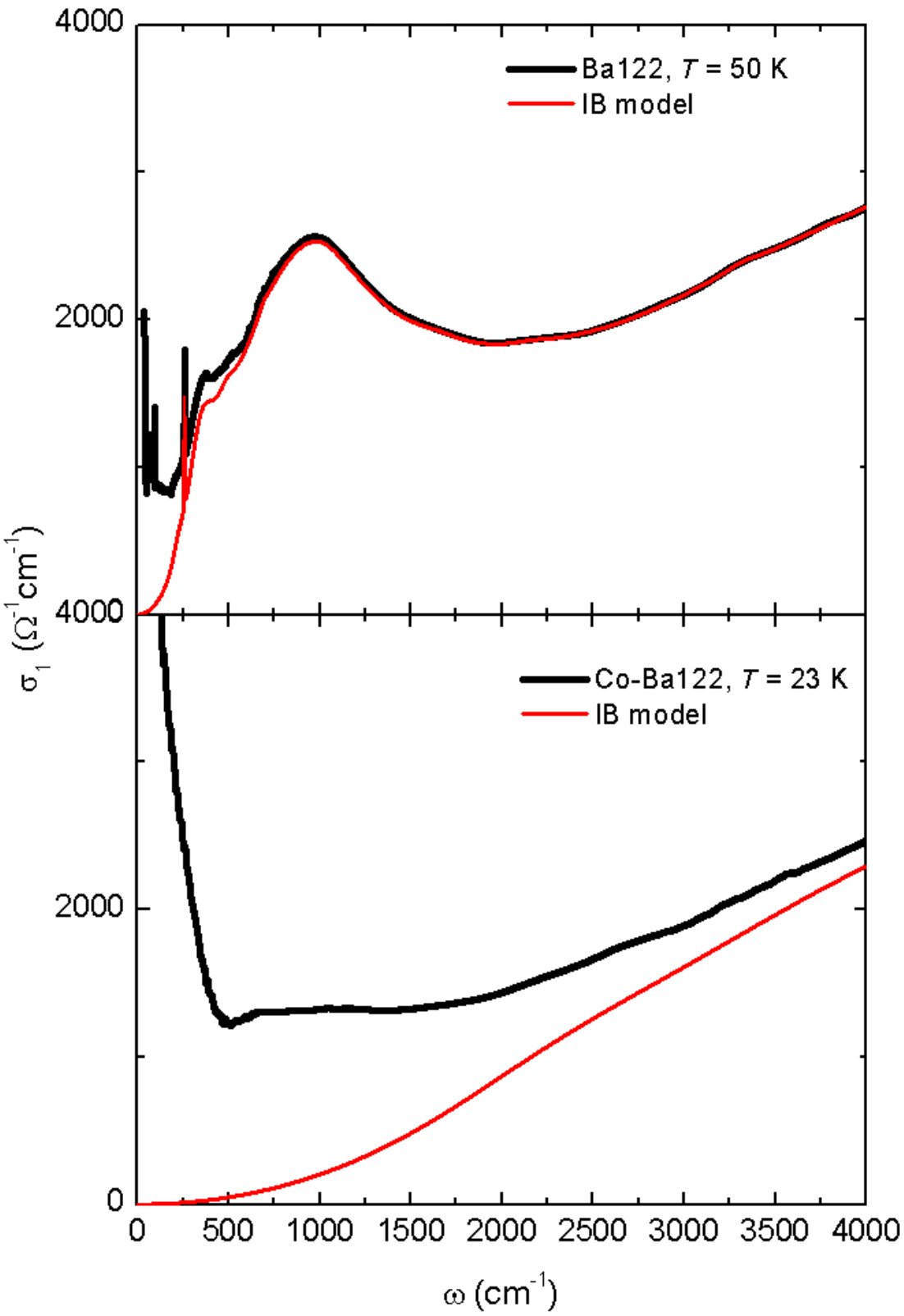}
\caption{Real part of the optical conductivity $\sigma_1$ at \emph{T} = 23K in Co-Ba122 with the modeled interband contribution. The IB model was subtracted from the $\sigma_1$ data in order to determine $\sigma_1$ due solely to the Drude response.}
\end{figure}

In table I, we show the values of the Drude plasma frequency $\omega_p$, ($K_{exp}(\Omega=\Omega_D)=\omega_p^2$), and the corresponding number designation in Figure 4. We note that in order to determine one value for $\omega_p$ for each method in both the PM and SDW state, we averaged the values across all temperatures corresponding with that state. (We did not, however, include the \emph{T} = 100K value in the SDW average.)

\begin{table}
\caption{$\omega_p$ values used for Figure 4, with the corresponding method we employed. $\omega_p$ under labels 1, 2, and 5 were determined by integrating the optical conductivity up to listed cut-off value, while values under labels 3 and 4 were determined by subtracting the listed IB model from the optical conductivity and integrating the remaining spectra.}
\begin{center}
\begin{tabular}{c | c | c | c | c }
\cline{1-5}
\multicolumn{5}{| c |}{$\omega_p$ in the PM State} \\
\cline{1-5}
\multicolumn{1}{| c ||}{Label} & \multicolumn{1}{| c |}{1} & \multicolumn{1}{| c |}{2} & \multicolumn{1}{| c |}{3} & \multicolumn{1}{| c |}{4} \\
\cline{1-5}
\multicolumn{1}{| c ||}{Method} & \multicolumn{1}{| c |}{$\sigma_1^{min}$} & \multicolumn{1}{| c |}{$\sigma_2$ = 0} & \multicolumn{1}{| c |}{IB model} & \multicolumn{1}{| c |}{$E_g^1$} \\
\cline{1-5}
\multicolumn{1}{| c ||}{Ba122} & \multicolumn{1}{| c |}{11,297} & \multicolumn{1}{| c |}{10,544} & \multicolumn{1}{| c |}{-} & \multicolumn{1}{| c |}{-} \\
\cline{1-5}
\multicolumn{1}{| c ||}{Co-Ba122} & \multicolumn{1}{| c |}{10,600} & \multicolumn{1}{| c |}{10,881} & \multicolumn{1}{| c |}{11,788} & \multicolumn{1}{| c |}{-} \\
\cline{1-5}
\end{tabular}
\end{center}

\begin{center}
\begin{tabular}{c | c | c | c | c }
\cline{1-5}
\multicolumn{5}{| c |}{$\omega_p$ in the SDW State} \\
\cline{1-5}
\multicolumn{1}{| c ||}{Label} & \multicolumn{1}{| c |}{1} & \multicolumn{1}{| c |}{2} & \multicolumn{1}{| c |}{3} & \multicolumn{1}{| c |}{4} \\
\cline{1-5}
\multicolumn{1}{| c ||}{Method} & \multicolumn{1}{| c |}{$\sigma_1^{min}$} & \multicolumn{1}{| c |}{$\sigma_2$ = 0} & \multicolumn{1}{| c |}{IB model} & \multicolumn{1}{| c |}{$E_g^1$} \\
\cline{1-5}
\multicolumn{1}{| c ||}{Ba122} & \multicolumn{1}{| c |}{3,130} & \multicolumn{1}{| c |}{3,628} & \multicolumn{1}{| c |}{3,414} & \multicolumn{1}{| c |}{4,060} \\
\cline{1-5}
\end{tabular}
\end{center}
\end{table}

Finally, we note that these results slightly underestimate $K_{exp}$ for two reasons. First, our integrals begin at a small but finite frequency, integrating primarily from the low-energy cut-off of real data points. In our case, we began the integrals in the PM state at 30 cm$^{-1}$ and in the SDW state at 10 cm$^{-1}$. This may underestimate $\omega_p$ by up to 10\% in the SDW state and up to 6\% in the PM state. The same error can be assumed to apply to $\omega_p$ determined from the IB models. Second, the cutoff points of our integrals in the PM state most likely do not completely account for the long decay of the incoherent Drude, which extends into the IB region. Overall, the values of $\omega_p$ in table I likely underestimate the actual $\omega_p$ by about 10\%. Therefore, we estimate the value of $K_{exp}/K_{LDA} \approx 0.4$ to be an upper bound of the kinetic energy ratio.

\section{C. Comparison with \emph{ARPES}}

Results from angle resolved photoemission spectroscopy (ARPES) experiments also require renormalization of band theory values \protect\cite{DingARPES,Yi-PRB80-024515-2009,Yi-arxiv09090831}. The kinetic energy ratio defined for ARPES is then based on the band renormalization factor required to make band theory predictions match the measured dispersion near and at the Fermi level. (We note that this factor may be different for energies far from the Fermi level). In the PM state of Ba122 and Co-Ba122 (\emph{x} = 0.06), bandwidth renormalization factors of 1.5 and 1.4, respectively, are necessary \protect\cite{Yi-PRB80-024515-2009}, corresponding to $K_{ARPES}/K_{LDA}$ = 0.66-0.71. In the SDW state of Ba122, ARPES finds the LDA renormalization factor doubles to $\approx$3 \protect\cite{Yi-arxiv09090831}, corresponding to $K_{ARPES}/K_{LDA}$ = 0.33. In the SDW state, using the integration cut-off value equal to the lower SDW gap energy $E_{g}^{1}$ = 336 cm$^{-1}$, we find $K_{exp}/K_{LDA}$ = 0.34, in agreement with the ARPES results. However, in order for optics to reproduce the ARPES results in the PM state, the integral would need to extend well into the interband region ($\Omega$= 3500 cm$^{-1}$) and would no longer represent an accurate estimate of the kinetic energy. Therefore, we are faced with a disagreement between ARPES and optics in the PM state.

The discrepancy between the two probes may be due to the difference in single-particle (ARPES) and two-particle (optical) dynamics, similar to conclusions recently drawn to explain discrepancies between infrared Raman spectroscopy and ARPES in a series of overdoped cuprates \protect\cite{Prestel-arXiv1009.0213}. A second possibility is that in the PM state, the incoherent quasiparticles cause the bands measured by APRES to be very broad and diffuse. This could make an accurate estimate of the renormalization value via ARPES difficult because it is based upon a second derivative of the raw energy spectra. Indeed, the importance of an incoherent quasiparticle contribution in the PM state and the disappearance of incoherent quasiparticles in the SDW state is clear from our kinetic energy analysis as well as from results in the literature \protect\cite{Tu-arxiv1008.3098,Fang-PRB80-140508R-2009,Nakajima-PRB81-104528-2010,WuPRB81}. In yet a third possibility, the discrepancy may stem from the fact that twinning of crystals complicates the Fermi surface, and may increase the Fermi surface area as measured by ARPES, subsequently leading to the underestimation of correlations in the PM state \protect\cite{arxiv:1103.3329}. What is clear from both spectroscopic probes, however, is that there is a substantial renormalization of bands in the pnictides.


\begin{thebibliography}{99}
\bibitem{footnote} As temperature is decreased in the high-temperature metallic phase of V$_2$O$_3$, SW moves from high to low energy into the Drude response, over an energy scale associated with the Hubbard $U$. Upon entrance into the insulating phase at low temperatures, the Drude response disappears and the associated SW moves from low to high energy.
\end{thebibliography}
\end{document}